# Multiple super-resolution imaging from Mikaelian lens to generalized Maxwell's fish-eye lens

YANGYANG ZHOU AND HUANYANG CHEN*

*Institute of Electromagnetics and Acoustics and Department of Physics, Xiamen University, Xiamen 361005, China*

*kenyon@xmu.edu.cn

**ABSTRACT**
Super-resolution imaging is vital for optical applications, such as high capacity information transmission, real-time bio-molecular imaging and nanolithography. Technology and method of super-resolution imaging have attracted much attention. In recent years, different kinds of novel lens, from the superlens to the super-oscillatory lens, has been designed and fabricated to break out the diffraction limit. However, the effect of the super-resolution imaging in these lenses is not satisfactory due to intrinsic loss, aberration, bigger sidebands and so on. Moreover, these lenses also cannot realize multiple super-resolution imaging. In this research, we introduce the solid immersion mechanism to Mikaelian lens (ML) for multiple super-resolution imaging. The effect is robust and valid for broadband frequencies. Based on conformal transformation optics as a bridge linking the solid immersion ML and generalized Maxwell's fish-eye lens (GMFEL), we also discover the effect of multiple super-resolution imaging in the solid immersion GMFEL.

## 1. INTRODUCTION

The resolution of the conventional lens is inherently constrained by the diffraction limit, wherein the spatial information of features smaller than one-half of the wavelength is evanescent and cannot be transferred to the far field. Although near-field scanning optical microscope has achieved super-resolution by collecting the evanescent field in close proximity to the object [1], this serial technique suffers from the slow scanning speed and non-negligible near-field perturbation preventing its application in real-time imaging. A perfect lens [2], relying on negative index materials [3, 4] to restore evanescent wave at the imaging point, as a first step towards real-time imaging was proposed in 2000 by Pendry. Following the concept of the perfect lens, a series of superlenses [5-11] were fabricated to project the sub-diffraction-limited imaging at the near field of the lenses. Later, a hyperlens [12] was designed to far-field super-resolution imaging by metamaterials with hyperbolic dispersion supporting the propagating waves with very large spatial wave vector. Utilizing alternating metal-dielectric structure in a curved geometry and other metamaterial structures, optical hyperlenses were fabricated [13-15] to project the sub-diffraction-limited magnified imaging at the far-field of the lens. However, for these designs both the superlens and hyperlens, insurmountable manufacturing challenges and intrinsic losses from the lens are big obstacles to the applications.

Recently, a series of super-oscillation lenses (SOLs) [16-18] have been designed and fabricated to achieve super-resolution imaging without evanescent waves based on optical super-oscillation theory [19, 20]. From theoretical analysis, we can conclude that the resolution of SOL goes to infinity. However, with the resolution increasing, more growing sidebands will emerge at the focus spots, which will severely constrain the focusing efficiency. The working bandwidth of SOL is limited to single frequency or some discrete frequency points, which is another obstacle. On the other hand, solid immersion lens (SIL) [21, 22] as a tool to yield super-resolution imaging has been studied extensively through the application of high refractive-index (RI) solid material and specific geometric optical design [23-25]. Utilizing total internal reflection (TIR) happening at the interface to excite evanescent waves, the SIL improves the imaging resolution. So far, many types of SILs have been developed from conventional structures to novel metamaterials structures [26-28]. However, chromatic aberration confines the applications of SIL and is limited to narrow-band emission spectra.

Recently, gradient RI (GRIN) lens has been developed rapidly for its excellent capability to control the propagation of electromagnetic waves [29-33] and enable focusing or imaging [34-41]. Among these GRIN lenses, the Mikaelian lens (ML) as a self-focusing cylindrical medium was derived firstly by Mikaelian in 1951[42] and has drawn much attention, due to its property of self-imaging in geometrical optics [43, 44]. Many applications based on the Mikaelian lens were designed and fabricated from microwave frequency to optical frequency [43, 45, 46]. If we combine the self–imaging performance of ML with solid immersion mechanism, super-resolution imaging is highly anticipated for ML.

In this work, we first proposed a solid immersion Mikaelian lens (SIML) and proved it can achieve multiple super-resolution real-time imaging along the edge of the lens without the need for optical scanning or image reconstruction by numerical simulation. Through slightly modifying the RI profile of the ML, evanescent waves ignited at the air/lens interface contribute to the process of super-resolution imaging. We will find that SIML provides no chromatic aberration imaging. Later on, we used conformal transformation [47, 48] to design a modified solid immersion generalized Maxwell's fish-eye lens (SIGMFEL) based on SIML and verify performance of the modified SIGMFEL for multiple super-resolution imaging. Although a drain-assisted GMFEL[49] with extreme RI profile also can achieve subwavelength imaging, the drain located at the imaging position hinders the information collection of the target object, and at same time the information of the imaging includes both from the object and from the drain which the imaging is no longer an intrinsic property of GMFEL.
Different from the drain-assisted GMFEL, this modified SIGMFEL only alters the RI profile maintaining the intrinsic property of GMFEL and circumventing complex drain. The modified SIGMFEL is easier to realize due to its RI profile with a feasible range than drain-assisted GMFEL and conventional generalized Maxwell's fish-eye lens (GMFEL) [50, 51].

We will start from a semi-infinite SIML and analyze the super-resolution imaging performance of the SIML. Considering practical applications, a truncated SIML with finite size and RI distribution of feasible range was proposed and can maintain the super-resolution imaging performance to the level of semi-infinite SIML. Later on, utilizing conformal transformation, a modified SIGMFEL was designed for multiple super-resolution imaging based on the truncated SIML. For the truncated SIML and the modified SIGMFEL without extreme RI profile, we theoretically prove the validity of super-resolution imaging. It paves the way for fabrication and applications of SIML and SIGMFEL. All the imaging functionalities of the lenses will be confirmed by full-wave numerical simulation and raytracing simulation from commercial software COMSOL.

## 2. RESULTS AND DISCUSSION

Generally, the Mikaelian Lens (ML) [42] is a cylindrical lens, whose RI distribution satisfies $n_0 \text{sec} h(\beta r)$ decreasing gradually from the center to the edge in the radial direction, where $r$ is radial distance and $n_0 = \alpha \beta$ is the maximum refraction index along the symmetric axis, $\beta$ is the gradient coefficient which determines the focusing period of the lens $L = 2\pi/\beta$. For a two-dimensional case, the RI can be written as (assuming the symmetric axis is along the y-axis):

$$n = n_0 \sec h(\beta x)$$

of which $\alpha$ and $\beta$ conventionally set as 1 and the RI on the symmetric axis matching RI of the air background. In this case, Fig. 1(a) shows one-half of conventional ML (with $n_0$=1, $\beta$=1) located in the region of ($0 \leq x < \infty, 0 \leq y \leq 5L/2$) in the air background, and no reflection and evanescent wave emerge at lens/air interface. Light rays propagate along a sine-like path focusing and the RI profile of ML is shown in Fig. 1(a). Different from Fig. 1(a), Fig. 1(b) introduces impedance mismatching at the lens/air interface by changing $n_0$=3 and the evanescent wave is ignited at the interface. Due to TIR at the edge of the lens, light rays realize multiple focusing points along the edge of the lens as shown Fig. 1(b). By full-wave simulation, we analyze field intensity profile of the SIML and its performance of imaging at the frequency of 3GHz. Figures 1(c) and (d) show the field intensity profile of the conventional ML and SIML and the corresponding full width at half maximum (FWHM) of which a point source (line current) (located at x=0, y=L/2) excites a transverse electric (TE) cylindrical wave in the lens, respectively. In figures, the solid red curves represent normalized electric field intensity along the y-axis direction at air imaging plane and the related FWHMs are marked as well. Clearly, the corresponding FWHM decreases from 0.98λ to 0.39λ with $n_0$=1 increasing to 3. It reveals that the resolution of the SIML is below the diffraction limit 0.5λ and keeps the subwavelength resolution along the edge of the lens in multiple focusing points. Predictably, the ML can achieve super-resolution imaging.

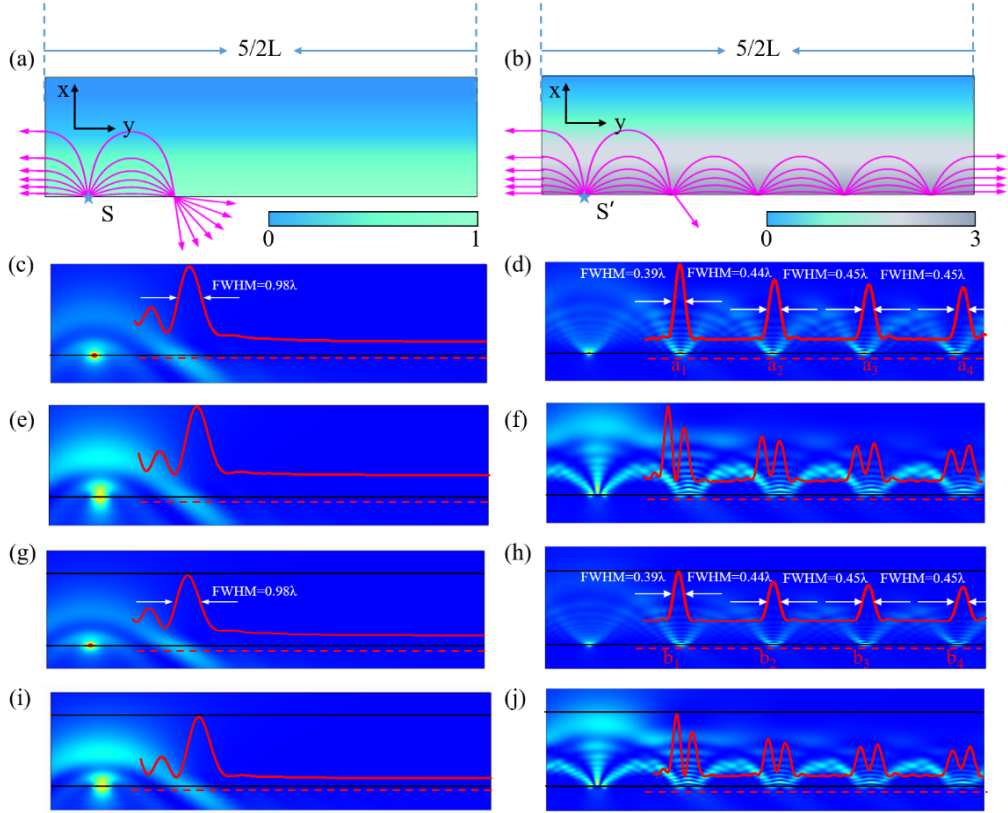

**Fig. 1.** Schematics and imaging functionalities of ML with $n_0=1$ (conventional ML) and SIML with $n_0=3$ respectively. Related results of conventional ML are shown in the left column of the figure and the right column present related results of SIML. (a) Schematic of a semi-infinite conventional ML with a gradient RI profile along x-axis direction and light trajectories in the lens. (b) Schematic of a semi-finite SIML with a gradient RI profile along x-axis direction and light ray trajectories in the lens. Multiple focusing points are formed along y-axis direction at the edge. (c, d) Calculated electric field intensity distributions and the corresponding FWHM of the semi-infinite conventional ML and SIML. The red curves present the normalized electric field intensity along y-axis direction distance 1mm from the bottom edge of the lens and the related FWHM is marked. (e, f) In the two semi-infinite ML, a pair of identical point sources with $1/3\lambda$ spacing was severed as excitation sources at the frequency of 3GHz and the electric field intensity distributions are shown respectively. Multiple super-resolution imaging emerges at air near the edge of the ML with $n_0=3$. (g, j) The truncated conventional ML and SIML with finite size and the related imaging functionalities. The black solid lines denote the boundary of the lens. The truncated SIML maintains the performance of super-resolution imaging consistent with the semi-finite SIML.

To further verify the super-resolution imaging of SIML, a pair of identical sources with $1/3\lambda$ spacing are located at the interface of the lens to excite a TE cylindrical wave as shown in Fig. 1(f). It shows that the SIML resolves the two-point sources and realizes super-imaging successfully. In the figure, the red solid curve illustrates the normalized field intensity along the y-axis direction from $L$ to $L/5$ and $x=-1$mm at the air. As comparison, the conventional ML interact with a pair of identical sources with $1/3\lambda$ spacing as shown in Fig. 1(c). According to the figure, the conventional ML fails to distinguish the two-point sources. Therefore, the SIML successfully realizes super-resolution imaging and the mechanism of TIR is valid for improving the resolution of the lens. However, for this semi-infinite SIML, it is difficult for fabrication and application. In addition, the RI profile of the lens along the x-axis direction gradually decreases to 0, which is another obstacle for practical fabrication. To circumvent the problem, we truncate the semi-infinite SIML into finite width and at the same time adjust RI ranging from 3 to 1 along the x-axis direction, as shown in Fig. 1(g) and (h). The truncated SIML is located in the region of ( $0 \leq x \leq \operatorname{arccosh}(3)$ , $0 \leq y \leq 5L/2$ ). Utilizing the full-wave simulation, we calculate the truncated SIML performance of super-resolution imaging, as shown in Fig. 1(h, j). Obviously, the functionality of the truncated SIML is not compromised to that of the semi-infinite SIML. It resolves the two identical sources with $1/3\lambda$ spacing and realizes imaging beyond the diffraction limit successfully. This paves a way to practical application. As a comparison, the imaging of the truncated conventional ML is also calculated by full-wave simulation, as shown in Fig. 1(g, i). The size of the truncated conventional ML is the same as that of the truncated SIML and the black solid line represents the boundary of the lens. In the figures, we can see the profile of electric field intensity and normalized electric field intensity at the imaging plane which distances 1mm from the bottom of the lens. It is clear that the truncated conventional ML fails to realize super-resolution imaging at air as well.

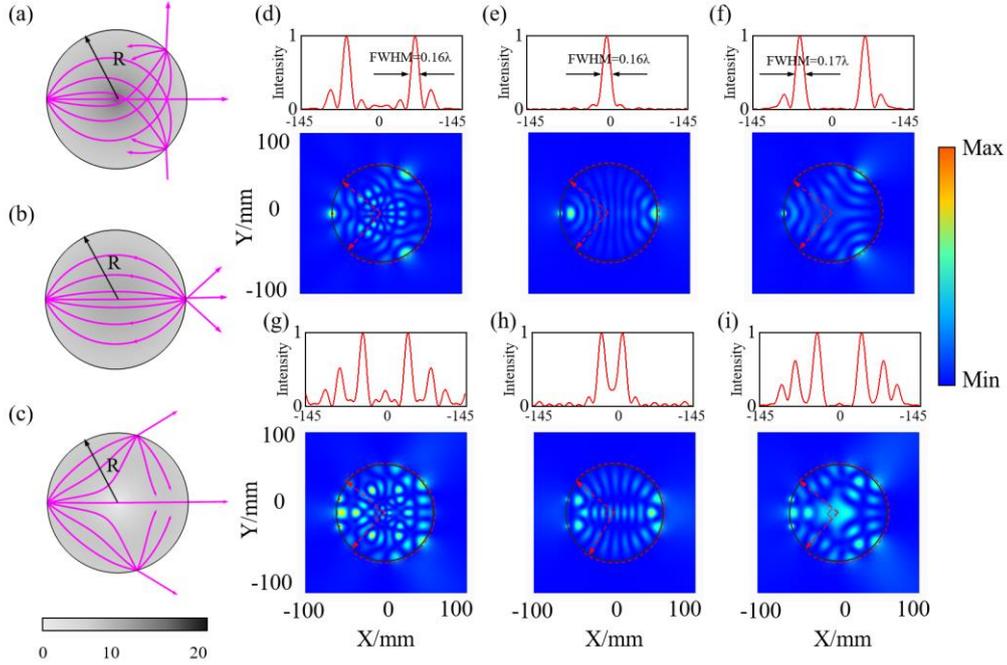

**Fig. 2.** Schematics and super-resolution imaging functionalities of the modified SIGMFELs with $\beta$ = 0.8, 1, and 1.7, respectively. (a, b, c) Schematic of the modified SIGMFELs with a gradient RI profile and light trajectories in the lenses for $\beta$ = 0.8, 1, and 1.7, respectively. All the rays emitting from the point source are converging at the edge of the lenses and a part of the light rays are reflected respectively. (d, e, f). Analysis of super-resolution imaging performance of the modified SIGMFLs with the value of $\beta$ = 0.8, 1, and 1.7 with a point source at the frequency of 3GHz respectively. The corresponding FWHM at three different SIMFELs are marked. (g, h, i) Imaging performance of the three modified SIGMFELs of which two points

sources with a spacing of 1/3λ are placed at the edge of the lens. The red curves present the normalized electric field intensity along a concentric arc with a radius 61mm of the lens from -135 degree to 135 degree at the air imaging plane and the related FWHMs are marked respectively.

We will use an exponential conformal mapping [47, 51] $w = \exp(-z(x,y))$ to design SIGMFEL based on the above semi-finite SIML with the RI profile $n = n_0 \sec h(\beta x)$ where $n_0$ =3. We can obtain the RI profile of the SIGMFEL in u-v space according to the following formula:

$$n_w = n_z \left| \frac{dz}{dw} \right| \quad (1)$$

By calculation, the RI profile of the designed SIGMFLE is $n = \frac{2n_0}{(r/R)^{1+\beta} + (r/R)^{1-\beta}}$, where $r$ is the distance from the center of the lens and $n_0$ represents ambient RI, and $R$ denotes the radius of SIGMFEL. Therefore, we derive the RI profile of the SIGMFEL. According to transformation optics, we can deduce that the SIGMFEL also achieves super-resolution imaging. To verify the super-resolution imaging performance of the SIGMFEL, we choose three different semi-finite SIMLs with $\beta$ = 0.8, 1, and 1.7 and transformed the three SIMLs into three circular SIGMFELs respectively, as shown in Fig. 2(a-c). The size of these SIGMFELs are same and the corresponding radius is 60mm. Light rays and the related RI profiles are shown in the figures. For $\beta$ = 0.8 and 1.7, all the rays emitting from the point source are converging at the two different points as present in Fig. 1(a, c), respectively. For $\beta$ =1, the GMFEL becomes well-known MFEL[40] and can focus all light rays from a point source into the opposite point and the RI decrease from 6 (at the center) to 3 (at the edge), as shown Fig. 2(b). A part of light rays are reflected at the edges of lenses due to impendence mismatching at the lens/air interfaces. The ambient RI of the three lenses is $n_0$=3. Next, we will stimulatingly calculate the super-resolution imaging of three different SIGMFEL at frequency 3GHz. A point source (line current) is located at x =−60mm, y=0mm (the center of the SIGMFEL is located at the origin) to excite a TE cylindrical wave. Figure 1(d-f) show the electric field intensity patterns and the corresponding FWHM of the three different GMFEL with $\beta$ = 0.8, 1, and 1.7 respectively. In the figures, the red curves display the electric field intensity at the imaging plane and the related FWHM of the imaging point in air. Notably, the corresponding FWHM is less than 0.2λ which is far below the diffraction limit 0.5λ at the frequency of 3GH for the three lenses. It demonstrates that the three GMFELs achieve super-imaging successfully.

To further verify the super-imaging of the lens, a pair of point sources with 1/3λ spacing are located at the edge of the lens as an excitation source at same frequency. From Fig. 2(g-i), the electric filed intensity are clearly shown. The related normalized electric field intensity of the air imaging plane, along a concentric arc with a radius 61mm of the lens from -135 degree to 135 degrees, are shown by the red solid curve. The four obvious peaks in the figures represent the imaging points. It is clear that the modified SIGMEL with $\beta$=0.8 and 1.7 can resolve the two sources with 1/3λ spacing and achieve multiple super-resolution imaging. For SIGMFEL with $\beta$=0.8 and 1.7, they can resolve the two source points and achieve multiply super-resolution imaging. For SIGMFEL with $\beta$ =1, it only achieves single super-resolution imaging.

Therefore, we prove that the three modified SIGMFEL can overcome the diffraction limit and realize super-resolution imaging. However, for the value of $\beta$ is not equal to 1, the RI profile of SIGMFEL at center tends to the extreme value, which is a big barrier to fabrication and application. For $\beta$ less than 1, the RI profile of the lens gradually increases to infinity from the edge to the center. For $\beta$ more than 1, the RI profile of the lens gradually decreases to 0 from the edge to the center. If the extreme RI profile of the lens can be adjusted within feasible range and at the same time the modified lens maintains its original functionality, the applications for super-resolution imaging will be highly anticipated.

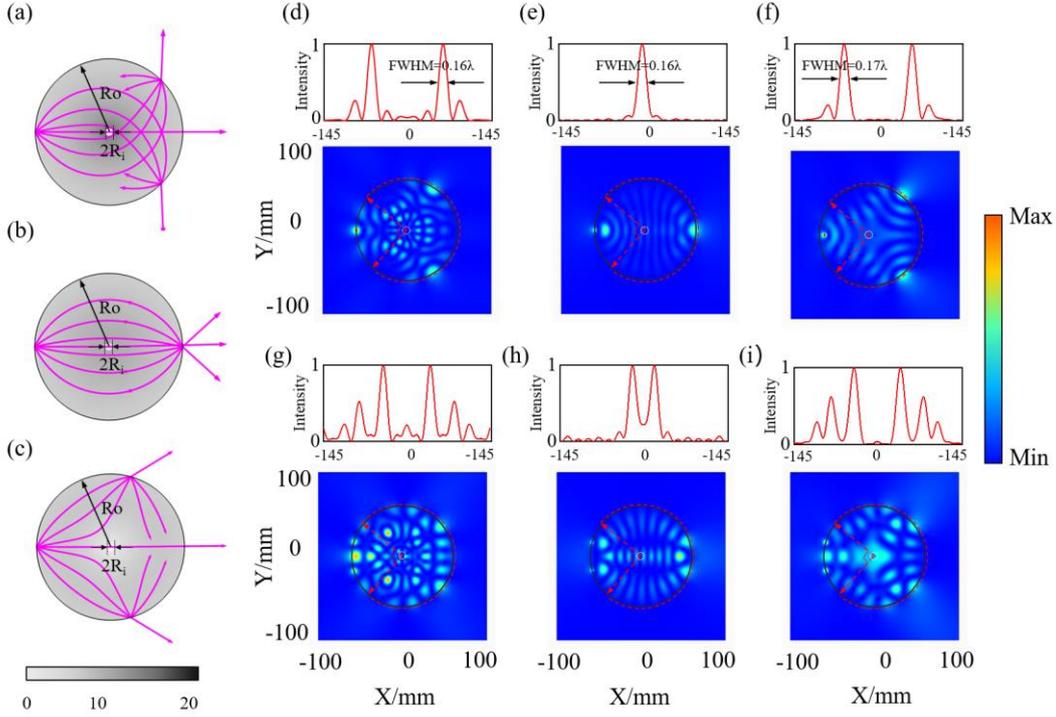

**Fig. 3.** Schematics and super-resolution imaging functionalities of the modified SIGMFELs with $\beta$ = 0.8, 1, and 1.7, respectively. (a, b, c) Schematic of the modified SIGMFELs with a gradient RI profile and light trajectories in the lenses for $\beta$ = 0.8, 1, and 1.7, respectively. All the rays emitting from the point source are converging at the edge of the lenses and a part of the light rays are reflected respectively. (d, e, f). Analysis of super-resolution imaging performance of the modified SIGMFLs with the value of $\beta$ = 0.8, 1, and 1.7 with a point source at the frequency of 3GHz respectively. The corresponding FWHM at three different SIMFELs are marked. It is far below the diffraction limit. (g, h, i) Imaging performance of the three modified SIGMFELs of which two points sources with a spacing of $1/3\lambda$ are placed at the edge of the lens. The red curves present the normalized electric field intensity along a concentric arc with a radius 61mm of the lens from -135 degree to 135 degree at the air imaging plane and the related FWHMs are marked respectively.

For the above problem, we propose the modified SIGMFEL without extreme RI profile and it can keep functionality of super-resolution imaging as same the original SIGMFEL. To design the modified SIGMFEL, we start with a optimized truncated SIML ($(0 \leq x \leq 4, 0 \leq y \leq 5L/2)$) with the RI profile $n = n_0 \operatorname{sech}(\beta x)$ where $n_0$=3 which can realize super-resolution imaging. Via utilizing the same exponential conformal transformation mapping $w = \exp(-z(x,y))$, based on the truncated SLML with $\beta$=0.8, 1, and 1.7, we obtain three annular SIGMFELs with outer radius $R_0$=60$mm$ and inner radius $R_i = \exp(-4)*R_0$ (as shown in Fig. 3(a-c). The relative RI distribution satisfies

$$n = \begin{cases} \dfrac{2n_0}{(r/R_o)^{1+\beta} + (r/R_o)^{1-\beta}}, & R_i \leq r \leq R_o \\ 1, & r \leq R_i \end{cases} \quad (2)$$

where $r$ is the distance from the center of the lens and $R_o$ denotes the outer radius and $n_0$=3 is ambient RI. Therefore, the extreme RI at the central region is removed from the lens. According to transformation optics, we can deduce that the three modified SIGMFELs will maintain the property of super-resolution imaging consistent with that of the truncated SIML.

To verify the performance of super-resolution imaging for the modified SIGMFEL with three values of $\beta$=0.8, 1, and 1.7, raytracing simulation and full-wave numerical simulation are performed at the frequency of 3 GHz, respectively. A point source is located at x=-60mm and y=0 to excite a cylindrical TE wave. Fig. 3(a-c) displays the imaging performance and reflection at the lens/ air interface for three value of $\beta$ = 0.8, 1, and 1.7 respectively. The figures also illustrate the RI distribution of modified SIGMFEL with three values of $\beta$=0.8, 1, and 1.7, respectively. Perfect geometric imaging and reflection happen at the lens/air interface. Figure 3(d-f) shows electric field intensity distribution in the three lenses and the related FWHMs are marked. It is clear that the corresponding FWHM is less than 0.2$\lambda$, which is far below the diffraction limit. The imaging performance of the modified SIGMFEL agrees well with the original GMFEL. This demonstrates that the modified SIGMMFEL can achieve super-resolution imaging as well. To further verify the super-imaging performance of the three modified SIGMFELs, full-wave numerical simulation is performance with a pair of identical point sources of 1/3$\lambda$ spacing at the frequency of 3GHz. The related electric field intensity distribution and normalized electric field intensity at the air imaging plane are shown in Fig. 3(g-i). Four tiny spots emerge at the edge of the two lenses with $\beta$=0.8 and 1.7 respectively in Fig. 3(g, i). The red solid curves present the normalized electric field intensity at the air imaging plane. The four obvious peaks in the figures represent the imaging points. It is clear that the modified SIGMEL with $\beta$=0.8 and 1.7 can resolve the two sources with 1/3$\lambda$ spacing and achieve multiple super-resolution imaging. The modified SIGMEFL with $\beta$=1 resolves the two sources with deep subwavelength spacing and the related normalized electric field intensity is shown by the red solid curve in Fig. 3(b). Therefore, we prove that the three modified SIGMFEL can overcome the diffraction limit and realize super-resolution imaging.

## 3. CONCLUSION
Enlightened by solid immersion lenses, we ingeniously introduce the TIR mechanism to excite evanescent waves at the lens/air interfaces for multiple super-resolution imaging of the SIML. Utilizing conformal mapping, we derive the RI profile SIGMFEL from the SIML. The SIML and the SIGMFEL could be used to overcome the diffraction limit. However, the extreme RI profile of the SIML and SIGMFEL have difficulties in fabrication and application. To circumvent the problem, we design a truncated SIML and a modified SIGMFEL without the extreme RI profile and verify the validity of the lenses for super-resolution imaging. In addition, the effect is robust and valid for broadband frequencies. It provides feasible designs for overcoming the diffraction limit from microwave to optical frequencies. It is possible to pave ways to multiple super-focusing, real-time bio-molecular imaging, nanolithography, high capacity information transmission waveguide, multichannel waveguide coupler, multichannel waveguide crossing.

**Funding.** National Natural Science Foundation of China (Grant No. 92050102); National Key Research and Development Program of China (Grant No. 2020YFA0710100); National Natural Science Foundation of China (Grant No. 11874311); Fundamental Research Funds for the Central Universities (Grant No. 20720200074).

**Disclosures.** The authors declare no conflicts of interest.

## References

1. B. Hecht, B. Sick, U. P. Wild, V. Deckert, R. Zenobi, O. J. F. Martin, and D. W. Pohl, "Scanning near-field optical microscopy with aperture probes: Fundamentals and applications," J. Chem. Phys. 112, 7761-7774 (2000).
2. J. B. Pendry, "Negative Refraction Makes a Perfect Lens," Phys. Rev. Lett. 85, 3966-3969 (2000).
3. V. Veselago, "The electrodynamics of substances with simultaneously negative values of permittivity and permeability," Sov. Phys. Usp. 10, 509 (1968).
4. R. A. Shelby, D. R. Smith, and S. Schultz, "Experimental verification of a negative index of refraction," Science **292**, 77-79 (2001).
5. N. Fang, H. Lee, C. Sun, and X. Zhang, "Sub-diffraction-limited optical imaging with a silver superlens," Science **308**, 534-537 (2005).
6. T. Taubner, D. Korobkin, Y. Urzhumov, G. Shvets, and R. Hillenbrand, "Near-field microscopy through a SiC superlens," Science **313**, 1595 (2006).
7. I. I. Smolyaninov, Y. J. Hung, and C. C. Davis, "Magnifying superlens in the visible frequency range," Science **315**, 1699-1701 (2007).
8. Z. W. Liu, S. Durant, H. Lee, Y. Pikus, N. Fang, Y. Xiong, C. Sun, and X. Zhang, "Far-field optical superlens," Nano Lett. **7**, 403-408 (2007).
9. X. Zhang, and Z. W. Liu, "Superlenses to overcome the diffraction limit," Nat. Mater. **7**, 435-441 (2008).
10. T. J. Huang, L. Z. Yin, J. Zhao, C. H. Du, and P. K. Liu, "Amplifying Evanescent Waves by Dispersion-Induced Plasmons: Defying the Materials Limitation of the Superlens," ACS Photonics **7**, 2173-2181 (2020).
11. Y. G. Bi, L. Peng, B. Yang, S. J. Ma, H. C. Chan, Y. J. Xiang, and S. Zhang, "Veselago lensing with Weyl metamaterials," Optica **8**, 249-254 (2021).
12. Z. Jacob, L. V. Alekseyev, and E. Narimanov, "Optical Hyperlens: Far-field imaging beyond the diffraction limit," Opt. Express **14**, 8247-8256 (2006).
13. Z. W. Liu, H. Lee, Y. Xiong, C. Sun, and X. Zhang, "Far-field optical hyperlens magnifying sub-diffraction-limited objects," Science **315**, 1686 (2007).
14. J. Rho, Z. L. Ye, Y. B. Xiong, X. Yin, Z. W. Liu, H. Choi, G. Bartal, and X. Zhang, "Spherical hyperlens for two-dimensional sub-diffractional imaging at visible frequencies," Nat. Commun. **1**, 1-5 (2010).
15. D. Lu, and Z. W. Liu, "Hyperlenses and metalenses for far-field super-resolution imaging," Nat. Commun. **3**, 1-9 (2012).
16. G. H. Yuan, E. T. Rogers, and N. I. Zheludev, "Achromatic super-oscillatory lenses with sub-wavelength focusing," Light Sci. Appl. **6**, e17036 (2017).
17. E. T. Rogers, J. Lindberg, T. Roy, S. Savo, J. E. Chad, M. R. Dennis, and N. I. Zheludev, "A super-oscillatory lens optical microscope for subwavelength imaging," Nat. Mater. **11**, 432-435 (2012).
18. Y. X. Shen, Y. G. Peng, F. Y. Cai, K. Huang, D. G. Zhao, C. W. Qiu, H. R. Zheng, and X. F. Zhu, "Ultrasonic super-oscillation wave-packets with an acoustic meta-lens," Nat. Commun. **10**, 1-7 (2019).
19. M. V. Berry, and S. Popescu, "Evolution of quantum superoscillations and optical superresolution without evanescent waves," J. Phys. A: Math. Gen. **39**, 6965-6977 (2006).
20. F. M. Huang, N. Zheludev, Y. Chen, and F. Javier Garcia de Abajo, "Focusing of light by a nanohole array," Appl. Phys. Lett. **90**, 091119 (2007).
21. S. M. Mansfield, and G. S. Kino, "Solid immersion microscope," Appl. Phys. Lett. **57**, 2615-2616 (1990).
22. D. A. Fletcher, K. B. Crozier, K. W. Guarini, S. C. Minne, G. S. Kino, C. F. Quate, and K. E. Goodson, "Microfabricated silicon solid immersion lens," J. Microelectromech. Syst. **10**, 450-459 (2001).
23. D. R. Mason, M. V. Jouravlev, and K. S. Kim, "Enhanced resolution beyond the Abbe diffraction limit with wavelength-scale solid immersion lenses," Opt. Lett. **35**, 2007-2009 (2010).
24. M. S. Kim, T. Scharf, M. T. Haq, W. Nakagawa, and H. P. Herzig, "Subwavelength-size solid immersion lens," Opt. Lett. **36**, 3930 (2011).
25. A. Bogucki, L. Zinkiewicz, M. Grzeszczyk, W. Pacuski, K. Nogajewski, T. Kazimierczuk, A. Rodek, J. Suffczynski, K. Watanabe, T. Taniguchi, P. Wasylczyk, M. Potemski, and P. Kossacki, "Ultra-long-working-distance spectroscopy of single nanostructures with aspherical solid immersion microlenses," Light Sci. Appl. **9**, 48 (2020).
26. H. Zhu, W. Fan, S. X. Zhou, M. Chen, and L. M. Wu, "Polymer Colloidal Sphere-Based Hybrid Solid Immersion Lens for Optical Super-resolution Imaging," ACS Nano **10**, 9755-9761 (2016).
27. A. Novitsky, T. Repän, R. Malureanu, O. Takayama, E. Shkondin, and A. V. Lavrinenko, "Search for superresolution in a metamaterial solid immersion lens," Phys. Rev. A **99**, 023835 (2019).
28. W. Fan, B. Yan, Z. B. Wang, and L. M. Wu, "Three-dimensional all-dielectric metamaterial solid immersion lens for subwavelength imaging at visible frequencies," Sci. Adv. **2**, e1600901 (2016).
29. Q. Wu, J. P. Turpin, and D. H. Werner, "Integrated photonic systems based on transformation optics enabled gradient index devices," Light: Science & Applications **1**, e38-e38 (2012).
30. C. He, J. T. Chang, Q. Hu, J. Y. Wang, J. Antonello, H. H. He, S. X. Liu, J. Y. Lin, B. Dai, D. S. Elson, P. Xi, H. Ma, and M. J. Booth, "Complex vectorial optics through gradient index lens cascades," Nat. Commun. **10**, 12286 (2019).
31. S. Li, Y. Y. Zhou, J. J. Dong, X. L. Zhang, E. Cassan, J. Hou, C. Y. Yang, S. P. Chen, D. S. Gao, and H. Y. Chen, "Universal multimode waveguide crossing based on transformation optics," Optica **5**, 1549-1556 (2018).
32. A. Forbes, "Common elements for uncommon light: vector beams with GRIN

lenses," Light Sci. Appl. **8**, 1-2 (2019).
33. M. G. Scopelliti, and M. Chamanzar, "Ultrasonically sculpted virtual relay lens for in situ microimaging," Light Sci. Appl. **8**, 65 (2019).
34. H. F. Ma, and T. J. Cui, "Three-dimensional broadband and broad-angle transformation-optics lens," Nat. Commun. **1**, 1126 (2010).
35. N. Kundtz, and D. R. Smith, "Extreme-angle broadband metamaterial lens," Nat. Mater. **9**, 129-132 (2010).
36. Y. Zhang, Y. He, H. W. Wang, L. Sun, and Y. K. Su, "Ultra-Broadband Mode Size Converter Using On-Chip Metamaterial-Based Luneburg Lens," ACS Photonics **10**, 0c01269 (2020).
37. U. Leonhardt, "Perfect imaging without negative refraction," New J. Phys. **11**, 093040 (2009).
38. U. Leonhardt, and T. G. Philbin, "Perfect imaging with positive refraction in three dimensions," Phys. Rev. A **81**, 011804 (2010).
39. S. C. Tao, Y. Y. Zhou, and H. Y. Chen, "Maxwell's fish-eye lenses under Schwartz-Christoffel mappings," Phys. Rev. A **99**, 013837 (2019).
40. Y. Zhou, Z. Hao, P. Zhao, and H. Chen, "Super-resolution imaging in absolute instruments," arXiv preprint arXiv:2107.01632 (2021).
41. T. Tyc, L. Herzánová, M. Šarbort, and K. Bering, "Absolute instruments and perfect imaging in geometrical optics," New J. Phys. **13**, 115004 (2011).
42. A. L. Mikaelian, and A. M. Prokhorov, "V Self-Focusing Media With Variable Index Of Refraction," in *Progress in Optics*(1980), pp. 279-345.
43. X. Y. Wang, H. Y. Chen, H. Liu, L. Xu, C. Sheng, and S. N. Zhu, "Self-Focusing and the Talbot Effect in Conformal Transformation Optics," Phys. Rev. Lett. **119**, 033902 (2017).
44. H. Y. Peng, S. L. Liu, Y. Z. Wu, Y. Yan, Z. C. Zhou, X. C. Li, Q. L. Bao, L. Xu, and H. Y. Chen, "Duplex Mikaelian and Duplex Maxwell's Fish-Eye Lenses," Phys. Rev. Appl. **13**, 034050 (2020).
45. V. V. Kotlyar, A. A. Kovalev, and V. A. Soifer, "Subwavelength focusing with a Mikaelian planar lens," Optical Memory and Neural Networks (Information Optics) **19**, 273-278 (2011).
46. F. Sun, Y. G. Ma, X. C. Ge, and S. L. He, "Super-thin Mikaelian's lens of small index as a beam compressor with an extremely high compression ratio," Opt. Express **2**, 7328-7336 (2013).
47. L. Xu, and H. Y. Chen, "Conformal transformation optics," Nat. Photonics **9**, 15-23 (2014).
48. U. Leonhardt, "Optical conformal mapping," Science **312**, 1777-1780 (2006).
49. Y. Yin, J. Li, and H. Chen, "Multiple drains in generalized Maxwell's fisheye lenses," Opt. Express **28**, 37218-37225 (2020).
50. R. K. Luneburg, *Mathematical theory of optics* (University of California Press, 1964).
51. H. Y. Chen, "Imaging along conformal curves," Phys. Rev. A **98**, 043843 (2018).